\documentclass[pra,aps,superbib,citeautoscript,twocolumn]{revtex4-1}

\usepackage[english]{babel}
\usepackage[utf8]{inputenc}
\usepackage[T1]{fontenc}

\usepackage[colorlinks=true,urlcolor=blue,citecolor=blue,linkcolor=blue]{hyperref}

\usepackage{amsmath,bm}
\usepackage{amstext}
\usepackage{epsfig}
\usepackage{xcolor}
\usepackage{subfig}
\usepackage{graphicx,amsmath,amssymb,tabularx}
\usepackage{multirow}
\usepackage{array}
\usepackage{dsfont}
\usepackage{caption}
\usepackage{cleveref}
\usepackage{ulem}
\usepackage[toc]{appendix}
\usepackage{physics}
\usepackage{mathrsfs}  

\definecolor{dgreen}{rgb}{0,.5,0}
\definecolor{dred}{rgb}{.7,.0,.0}

\def\ddroit{{d}}

\DeclareMathAlphabet\mathbfcal{OMS}{cmsy}{b}{n}
\newcommand{\bw}{\mathbf{w}}
\newcommand{\bxi}{\bm{\xi}}
\newcommand{\br}{\mathbf{r}}

\newcommand{\tw}{{\tt w}}

\newcommand{\be}{\begin{eqnarray}}
\newcommand{\ee}{\end{eqnarray}}
\DeclareMathAlphabet\mathbfcal{OMS}{cmsy}{b}{n}

\begin{document}

\title{
Individual correlations in ensemble density-functional theory:
State-driven/density-driven decompositions without additional Kohn--Sham systems
}

\author{
Emmanuel Fromager$^1$
}

\affiliation{\it 
~\\
$^1$Laboratoire de Chimie Quantique,\\
Institut de Chimie, CNRS / Universit\'{e} de Strasbourg,\\
4 rue Blaise Pascal, 67000 Strasbourg, France\\
}


\begin{abstract}
Gould and Pittalis [Phys. Rev. Lett. 123, 016401 (2019)] recently revealed a density-driven (DD) correlation energy that is specific to many-electron ensembles and must be accounted for by approximations. We derive in this Letter a general and simpler expression in terms of the ensemble weights, the ensemble Kohn-Sham (KS) orbitals, and their linear response to variations in the ensemble weights. As no additional state-driven KS systems are needed, its evaluation is greatly simplified. We confirm the importance of DD effects and introduce a direct and promising route to approximations.
\end{abstract}

\maketitle





{\it Introduction.} Time-dependent density-functional
theory (TD-DFT)~\cite{runge1984density} has become over the last two
decades the method of choice
for modeling properties of electronically-excited molecules and
materials. Despite this success, which is explained by the moderate
computational cost of the method, TD-DFT still suffers from various
deficiencies. The latter drastically reduce its applicability, in particular
to a large variety of molecules and materials where electron correlation is strong~\cite{Casida_tddft_review_2012}.
These failures originate from the single-reference
perturbative character of the theory [in the widely used linear response
regime] {and}
the common adiabatic approximation, where memory effects are absent. 
As a result, the interest in {\it time-independent}
formulations of DFT for excited
states has increased substantially over the last
decade~\cite{ayers2009pra,JCP09_Ziegler_relation_TD-DFT_VDFT,JCTC13_Ziegler_SCF-CV-DFT,PRA13_Pernal_srEDFT,franck2014generalised,yang2014exact,pribramjones2014excitations,pernal2015excitation,yang2017direct,gould2017hartree,gould2018charge,deur2017exact,deur2018exploring,sagredo2018can,senjean2018unified,nikiforov2014assessment,filatov2015spin,filatov2015ensemble,filatov2016self,filatov2017description,levy2016computation,deur2019ground}.\\
Gross--Oliveira--Kohn (GOK) ensemble
DFT~\cite{gross1988rayleigh,gross1988density,oliveira1988density}, which
is a generalization of Theophilou's DFT for
equiensembles~\cite{JPC79_Theophilou_equi-ensembles,theophilou_book}, is
one of these (low-cost) alternatives. Unlike state-averaged 
quantum chemical methods~\cite{helgaker2014molecular}, GOK-DFT 
describes (in principle exactly)
each state that belong to the ensemble with a single Slater determinant
(or a configuration state function),
in analogy with regular ground-state Kohn--Sham (KS) DFT.  
A substantial
difference with the latter though is that, in GOK-DFT, the
non-interacting KS ensemble is expected to reproduce the true
interacting ensemble
density [i.e. the weighted sum of ground- and excited-state densities]
only,
not each individual (ground- or excited-state) density. This subtle
point, which has not been much emphasized in the literature until very
recently~\cite{PRL19_Gould_DD_correlation,Gould_2020_insights}, is
central in the Gould-Pittalis correlation energy decomposition into  
state-driven (SD) and 
density-driven (DD) contributions~\cite{PRL19_Gould_DD_correlation}.\\  
This decomposition shed a new light on individual correlations
within an
ensemble and is relevant to the design of density-functional
approximations for ensembles,
which is an important outstanding problem in DFT~\cite{yang2017direct,deur2019ground,senjean2018unified,senjean2019ncentered}.
The way such a decomposition should be written and implemented is, however, open to discussion. Gould and
Pittalis~\cite{PRL19_Gould_DD_correlation} proposed to introduce
state-specific KS systems (one for each state, in
addition to the KS ensemble) which are expected to reproduce the
exact individual-state densities. For real {\it ab initio} (and
therefore {\it Coulombic})
systems, a unique KS system can indeed be
designed from a given excited-state density by 
requiring, as an additional constraint, the KS ground-state density to be as close as possible to
the ground-state Coulomb one, as shown by Ayers {\it et
al.}~\cite{PRA12_Nagy_TinD-DFT_ES,JCP15_Ayers_KS-DFT_excit-states_Coulomb,Ayers2018_Article_Time-independentDensityFunctio}
If model systems are considered instead (in order to minimize numerical
efforts), one may then face non-uniqueness or
representability issues when constructing a KS potential for each
excited state within the ensemble. The non-uniqueness 
problem can be solved through a selection
procedure~\cite{PRL19_Gould_DD_correlation}. But there might also be
situations where such a potential does
not exist. 
A simple 
example is
given by the two-electron asymmetric Hubbard dimer~\cite{carrascal2015hubbard}
where the occupation of the atomic sites plays the role of the density. In the non-interacting dimer, the density of the first singlet
excited state does not vary with the KS potential. It matches the
interacting excited-state density only when the dimer is symmetric~\cite{deur2017exact}.\\ 
As shown in this Letter, the non-uniqueness or non-existence of
excited-state KS potentials is not a problem as such in the context of 
ensemble DFT, where the KS potential is well defined (up to a
constant)~\cite{gross1988density}, simply because individual-state
properties can be extracted
in principle exactly from the KS density-functional ensemble. 
An exact expression for the individual density-functional correlation energies
is derived and, on that basis, two different  
SD/DD decompositions are explored. While being simpler and more general 
(i.e. applicable to all systems) than the Gould--Pittalis
decomposition~\cite{PRL19_Gould_DD_correlation}, one of them offers a
clearer way to model ensemble correlation energies. 

{\it A brief review of GOK-DFT.} Let us consider the $M+1$ lowest (in
energy)
solutions
to the electronic Schr\"{o}dinger equation $\hat{H}\Psi_I=E_I\Psi_I$,
$0\leq I\leq M$, 
where
the Hamiltonian 
$\hat{H}=\hat{T}
+\hat{W}_{\rm ee}+\hat{V}_{\rm ext}$ is the sum of the $N$-electron
kinetic energy, Coulomb repulsion, and local multiplicative external potential $\hat{V}_{\rm
ext}\equiv\sum^N_{i=1}v_{\rm ext}(\br_i)\times$ operators,
respectively.  
For simplicity, we will assume that the energies are not degenerate, i.e.
$E_0<E_1<\ldots<E_M$. Note that the theory can be easily extended to 
multiplets by assigning the same ensemble weight to degenerate
states~\cite{gross1988density}. The ensemble energy $E^{\mathbf{w}}=\sum^M_{I=0}{\tt
w}_IE_I$
is a weighted sum of ground- and excited-state energies where the
(positive) ensemble weights
decrease with increasing index $I$. They are normalized, i.e.
${\tt w}_0=1-\sum^{M}_{I=1}{\tt w}_I$, so that only the weights assigned to the excited states 
$\mathbf{w}\equiv ({\tt w}_1,{\tt
w}_2,\ldots,{\tt
w}_M)$ 
are
allowed to vary {\it independently}.\\
In GOK-DFT, the ensemble energy is determined as
follows for given and fixed weights
$\bw$~\cite{gross1988density}:
\be\label{eq:min_gamma_Ew}
E^{\mathbf{w}} =\min_{\left\{\varphi_k\right\}}
\left\{{\rm Tr}\left[\hat{\gamma}^\bw\left(\hat{T}
+\hat{V}_{\rm ext}\right)\right]
+E_{\rm
Hxc}^{\mathbf{w}}\left[n_{\hat{\gamma}^\bw}\right]\right\},
\ee
where ${\rm Tr}$ denotes the trace,
$\hat{\gamma}^\bw=\sum^M_{I=0}\tw_I\ket{\Phi_I}\bra{\Phi_I}$, and
$n_{\hat{\gamma}^\bw}(\br)
=
\sum^M_{I=0}\tw_In_{\Phi_I}(\br)$ is a trial ensemble density. The trial
determinants (or configuration state functions) $\Phi_I$ are all generated from the same set
$\left\{\varphi_k\right\}$ of orthonormal molecular orbitals that are optimized
variationally. The ensemble Hartree-exchange-correlation (Hxc) density functional in
Eq.~(\ref{eq:min_gamma_Ew}) can be decomposed exactly as 
$E^{\mathbf{w}}_{\rm Hxc}[n]
=
E^{\mathbf{w}}_{\rm Hx}[n]
+
E^{\mathbf{w}}_{\rm c}[n],
$
where the ensemble 
density-functional Hx energy~\cite{gould2017hartree}
\be\label{eq:EEXX}
E^{{\bf w}}_{\rm Hx}\left[n\right]=\sum^M_{K=0}{\tt
w}_K\;\bra{\Phi^{\bw}_K[n]}\hat{W}_{\rm ee}\ket{\Phi^{\bw}_K[n]}
\ee
is evaluated from the KS ensemble that reproduces the density $n$:
\be\label{ens_dens_KScontraint}
\sum^M_{K=0}{\tt
w}_K\;n_{\Phi_K^{\bw}[n]}({\bf r})
=n(\br).
\ee
Note that, in the general case, the $N$-electron KS wavefunctions
$\left\{\Phi^{\bw}_K[n]\right\}_{0\leq K\leq M}$ can be configuration
state functions~\cite{gould2017hartree}, i.e. linear combinations of KS
determinants. They are in principle
weight-dependent so that the 
density $n$ can be reproduced, whatever the value of the ensemble
weights~\cite{franck2014generalised,deur2017exact}. The minimizing KS
wavefunctions $\left\{\Phi^{\bf
w}_I=\Phi^{\bf w}_I\left[n^\bw\right]\right\}_{0\leq I\leq M}$ in
Eq.~(\ref{eq:min_gamma_Ew}) reproduce the exact ensemble density $n^{{\bf w}}$: 
\be\label{eq:phys_ens_dens_from_KS_ens}
\sum^{M}_{I=0}{\tt w}_In_{\Phi^{\bf w}_I}(\br)
=\sum^{M}_{I=0}{\tt
w}_In_{\Psi_I}(\br)
= n^{{\bf w}}({\bf r}),
\ee
so that the exact ensemble energy can be expressed as
\be\label{eq:final_exp_Ew}
E^{\bw}=\sum^M_{I=0}{\tt w}_I\bra{\Phi^{\bf w}_I}\hat{T}
+\hat{V}_{\rm ext}
\ket{\Phi^{\bf w}_I}+E_{\rm Hxc}^{\mathbf{w}}[n^{\bw}].
\ee
The corresponding minimizing orbitals fulfill the ensemble KS
equations~\cite{gross1988density},
\be\label{eq:ens_KS_sc_eq}
\left[-\dfrac{\nabla^2}{2}+
v_{\rm ext}(\br)+v^\bw_{\rm
Hxc}\left[n^\bw\right](\br)
\right]\varphi^\bw_p(\br)=\varepsilon^\bw_p\varphi^\bw_p(\br),
\ee
where
$v^\bw_{\rm
Hxc}\left[n\right](\br)=\delta E_{\rm Hxc}^{\mathbf{w}}[n
]/\delta n(\mathbf{r})$ is the ensemble Hxc density-functional potential.
When the KS wavefunctions are single determinants, their density simply
reads 
\be\label{eq:det_dens_from_KS_orbs}
n_{\Phi^{\bf w}_I}(\br)=\sum_{p
}\theta_p^I\left\vert\varphi^\bw_p(\br)\right\vert^2
,
\ee
where $\theta_p^I$ is the
(fixed and integer) occupation number of the orbital $\varphi^\bw_p$ in the determinant
$\Phi^{\bf w}_I$.
\\

{\it Extracting exact individual densities.} As pointed out in
Ref.~\cite{PRL19_Gould_DD_correlation}, Eq.~(\ref{eq:phys_ens_dens_from_KS_ens}) does not imply
that the KS wavefunctions reproduce the exact individual densities
$\left\{n_{\Psi_I}\right\}_{0\leq I\leq M}$. Nevertheless, these densities can be extracted directly from the KS
ensemble, as we will see. This means that it is in principle not
necessary to refer to additional
state-specific KS systems for modeling individual-state properties
within an ensemble.\\ 
We start from the simple observation that, like the energy~\cite{deur2019ground}, the density of any (ground or
excited) state
can be extracted from the
(linear-in-$\bw$) ensemble density as follows: 
\be\label{eq:phys_dens_from_ens_dens}
{n}_{\Psi_{{J}}}({\bf r})&=&
n^{{\bf w}}({\bf
r})+\sum^{M}_{I=1}\left(\delta_{I{J}}-{{\tt w}_I}\right)\dfrac{\partial
n^{{\bf w}}({\bf r})}{\partial {{\tt w}_I}}.
\ee
By inserting the KS ensemble density expression of 
Eq.~(\ref{eq:phys_ens_dens_from_KS_ens}) into
Eq.~(\ref{eq:phys_dens_from_ens_dens}) we obtain our first key result,
namely the exact
deviation in density of the true interacting state from the KS one, 
\be\label{eq:deviation_true-KS_dens}
{n}_{\Psi_{{J}}}({\bf r})
-n_{\Phi^\bw_J}({\bf r})&=&
\sum^{M}_{{I}=1}\sum^{M}_{K=0}\left(\delta_{I{J}}-{{\tt
w}_I}\right){\tt w}_K\dfrac{\partial n_{{\Phi}_K^{{\bf w}}}({\bf
r})}{\partial {{\tt w}_I}},
\nonumber\\
\ee
where $\partial n_{{\Phi}_K^{{\bf
w}}}(\br)/\partial\tw_I=2\sum_p\theta_p^K\;\varphi^\bw_p(\br){\partial
\varphi^\bw_p(\br)}/{\partial{\tt w}_I}$. As readily seen, the (static) linear response of the KS orbitals 
to variations in the ensemble
weights becomes central. In practice, it may be
evaluated by finite differences.
A numerically more robust approach, which requires additional implementation
work, would consist in solving an ensemble coupled-perturbed equation
which is 
derived in the supplementary material~\cite{supp_mat}, by analogy with
Ref.~\cite{JCP17_Filatov_REKS_indiv_energies_deriv}.\\ 

{\it Individual Hxc energies.} The next step 
consists in extracting individual
Hxc density-functional energies from the KS ensemble. For that purpose, we use the analog of Eq.~(\ref{eq:phys_dens_from_ens_dens}) for
energies~\cite{deur2019ground} which, when combined with
the {\it variational} KS expression of the ensemble energy in
Eqs.~(\ref{eq:min_gamma_Ew}) and (\ref{eq:final_exp_Ew}), leads to the
following exact energy level expression:    
\be\label{eq:ind_ener_from_KS}
E_J=\bra{\Phi_J^{\bw}}\hat{T}+\hat{V}_{\rm
ext}\ket{\Phi_J^{\bw}}+E^{{\bf w}}_{{\rm Hxc},{J}}\left[n^{{\bf w}}\right]
,
\ee
where the individual density-functional Hxc energy reads   
\be\label{eq:Hxc_J}
&&E^{{\bf w}}_{{\rm Hxc},{J}}\left[n\right]=
E^{{\bf w}}_{\rm Hxc}[n]+
\sum^{M}_{I=1}\left(\delta_{IJ}-{{\tt w}_I}\right)
\dfrac{{\partial E^{{\bf w}}_{\rm
Hxc}[}n{]}}{{\partial}{ {\tt w}_I}}
\nonumber\\
&&+
\int\ddroit{\bf r}\;
\dfrac{\delta {E^{{\bf w}}_{\rm Hxc}[}{n}
{]}}{\delta n({\bf r})}
\Big(n_{\Phi_J^{\bw}[n]}({\bf r})-{n}({\bf r})\Big).
\ee
Note that, as expected, the
ensemble density-functional Hxc energy is recovered from the weighted sum of the individual
Hxc energies [see Eqs.~(\ref{ens_dens_KScontraint}) and
(\ref{eq:Hxc_J})]:
\be\label{eq:sum_Jw_JHxc}
\displaystyle\sum^{M}_{{J}=0}{\tw_{{J}}}
\;
E^{{\bf w}}_{{\rm Hxc},{J}}\left[n\right]
=E^{{\bf w}}_{\rm
Hxc}[n].
\ee
Eqs.~(\ref{eq:Hxc_J}) and (\ref{eq:sum_Jw_JHxc}) establish a clearer connection
between ensemble and individual density-functional Hxc energies.  
Before analyzing the Hx and correlation terms separately for each state,
it is worth noticing that, according to
Eqs.~(\ref{eq:min_gamma_Ew}), (\ref{eq:phys_ens_dens_from_KS_ens}), and
(\ref{eq:final_exp_Ew}), the individual Hxc energy can be rewritten as
follows:  
\be\label{eq:ind_Hxc_ener-dallnw}
&&E^{{\bf w}}_{{\rm Hxc},{J}}\left[n^{{\bf w}}\right]
=
E^{{\bf w}}_{\rm
Hxc}\left[n^{{\bf w}}\right]
+\sum^{M}_{I=1}\left(\delta_{IJ}-{{\tt w}_I}\right)
\nonumber\\
&&
\times\left[
\dfrac{\partial
}{{\partial}{ {\tt w}_I}}
\Bigg(
E^{{\bf w}}_{\rm
Hxc}[n^{{\bf w}}{]}
\Bigg)-\left.\dfrac{\partial E_{\rm Hxc}^{\bxi}\left[n^{\bxi,                 \bw}\right]}{\partial{\tt w}_I}\right|_{\bxi=\bw}
\right],
\ee
where the auxiliary double-weight ensemble KS density
\be\label{eq:two_weight-ens_dens}
n^{\bxi,\bw}(\br)=\sum^M_{K=0}\xi_K\;n_{\Phi^{\bw}_K}(\br)
\ee
has been introduced. The term that is subtracted on the right-hand side of
Eq.~(\ref{eq:ind_Hxc_ener-dallnw}) originates from the
fact that the ensemble energy is calculated variationally. It is in
principle nonzero since the individual densities in the KS ensemble are {weight-dependent}, unlike in the true physical
system.\\

{\it Exact individual Hartree-exchange energies.} Let us first focus on
the individual Hx contributions to Eq.~(\ref{eq:ind_Hxc_ener-dallnw}).
As the dependence in $\bxi$ of the double-weight ensemble density in
Eq.~(\ref{eq:two_weight-ens_dens}) does not affect the individual KS
densities, we conclude that $\Phi^{\bxi}_K[n^{\bxi,\bw}]=\Phi^\bw_K$,
thus leading to [see Eq.~(\ref{eq:EEXX})],
\be
E_{\rm
Hx}^{\bxi}\left[n^{\bxi,\bw}\right]=\sum^M_{K=0}\xi_K\;\bra{\Phi^{\bw}_K}\hat{W}_{\rm
ee}\ket{\Phi^{\bw}_K},
\ee
while $E^{{\bf w}}_{\rm Hx}\left[n^\bw\right]=\sum^M_{K=0}\tw_K\;\bra{\Phi^{\bw}_K}\hat{W}_{\rm  ee}\ket{\Phi^{\bw}_K}$. As a result, the individual Hx
energy in Eq.~(\ref{eq:ind_Hxc_ener-dallnw}) reduces to the simple and
intuitive expression:
\be
E^{{\bf w}}_{{\rm Hx},{J}}\left[n^{{\bf
w}}\right]=\bra{\Phi^{\bw}_J}\hat{W}_{\rm ee}\ket{\Phi^{\bw}_J},
\ee
where, as emphasized previously, ${\Phi^{\bw}_J}$ can be a
configuration state function~\cite{gould2017hartree}.\\

{\it State- and density-driven correlations.}
We now focus on the individual correlation energies and their subsequent
SD/DD decomposition. We start
from the decomposition into individual components of the ensemble
density-functional correlation energy~\cite{PRL19_Gould_DD_correlation}, 
\be\label{eq:decomp_corr_ens_compt}
E_{\rm c}^\bw[n]=\sum^M_{K=0}\tw_K\;\mathcal{E}_{{\rm c},K}^\bw[n],
\ee
where $\mathcal{E}_{{\rm
c},K}^\bw\left[n^\bw\right]=\bra{\Psi_K}\hat{T}+\hat{W}_{\rm ee}\ket{\Psi_K}
-f^\bw_K\left[n^\bw\right]$ and
$f^\bw_K\left[n^\bw\right]=\bra{\Phi^{\bw}_K}\hat{T}+\hat{W}_{\rm
ee}\ket{\Phi^{\bw}_K}$ is the $K$th component of the 
exchange-only GOK functional $f^\bw[n]:=\sum^M_{K=0}\tw_Kf_K^\bw[n]$. By inserting
Eq.~(\ref{eq:decomp_corr_ens_compt}) into
Eq.~(\ref{eq:ind_Hxc_ener-dallnw}) and using
Eq.~(\ref{eq:deviation_true-KS_dens}) we obtain the following exact
expression for the individual correlation energy within the ensemble,   
\be\label{eq:exact_ind_corr_from_compt}
&&E^{{\bf w}}_{{\rm c},{J}}\left[n^{{\bf w}}\right]
=
\mathcal{E}_{{\rm
c},J}^\bw\left[n^\bw\right]
\nonumber
\\
&&
+\sum^M_{K=0}\tw_K\sum^{M}_{I=1}\left(\delta_{IJ}-{{\tt w}_I}\right)
\dfrac{\partial
}{\partial\tw_I}
\Big(\mathcal{E}_{{\rm
c},K}^{\bw}\left[n^{\bw}\right]
\Big)
\nonumber
\\
&&+\sum^M_{K=0}
\tw_K
\int
d\br\;\dfrac{\delta
\mathcal{E}_{{\rm
c},K}^{\bw}\left[n^{\bw}\right]
}{\delta
n(\br)}
\Big(n_{\Phi^\bw_{J}}(\br)-n_{\Psi_J}(\br)\Big),
\ee
which, as readily seen, is {\it not} equal to the $J$th component $\mathcal{E}_{{\rm
c},J}^\bw\left[n^\bw\right]$ of the
ensemble correlation energy. 
This is a major difference between exchange
and correlation energies in ensembles. As shown in the supplementary
material~\cite{supp_mat}, it originates from the fact that 
individual correlation energies incorporate
the density correction that must be applied to each KS state in order to
recover the exact individual external potential energies. 
\\ 

Interestingly, a first SD/DD
decomposition, that we may refer to as {\it density}-based, naturally emerges from
Eq.~(\ref{eq:exact_ind_corr_from_compt}). Indeed, substituting the true individual
densities for the KS ones leaves the first two terms on the right-hand
side unchanged (as the KS and true ensemble densities match) while the last term vanishes. Thus we may define
individual SD correlation energies as follows: 
\be\label{eq:1stdef_SD_corr}
\begin{split}
&E^{{\bf w},{\rm
SD}}_{{\rm c},{J}}\left[n^{{\bf w}}\right]:=
\mathcal{E}_{{\rm
c},J}^\bw\left[n^\bw\right]
\\
&
+\sum^M_{K=0}\tw_K\sum^{M}_{I=1}\left(\delta_{IJ}-{{\tt w}_I}\right)
\dfrac{\partial
}{\partial\tw_I}
\Big(\mathcal{E}_{{\rm
c},K}^{\bw}\left[n^{\bw}\right]
\Big),
\end{split}
\ee
or, equivalently,
$E^{{\bf w},{\rm
SD}}_{{\rm c},{J}}\left[n^{{\bf w}}\right]=
E^{{\bf w}}_{\rm
c}\left[n^{{\bf w}}\right]+
\sum^{M}_{I=1}\left(\delta_{IJ}-{{\tt w}_I}\right)
\frac{\partial
}{\partial{\tt w}_I}
\left(E^{{\bf w}}_{\rm
c}[n^{\bf w}]
\right)$.
As readily seen from Eq.~(\ref{eq:ind_Hxc_ener-dallnw}), the
complementary DD contribution (third term on the right-hand side of
Eq.~(\ref{eq:exact_ind_corr_from_compt})) will then be defined as  
\be\label{eq:alternative_exp_DD_corr}
E^{{\bf w},{\rm
DD}}_{{\rm c},{J}}\left[n^{{\bf w}}\right]
&:=&-\sum^{M}_{{I}=1}\left(\delta_{I{J}}-{{\tt
w}_I}\right)\left.\dfrac{\partial E_{\rm
c}^{\bxi}\left[n^{\bxi,\bw}\right]}{\partial{\tt
w}_I}\right|_{\bxi=\bw}.
\ee
Such a decomposition is of course arbitrary and not unique. One may opt
for a more {\it state}-based approach 
where only the terms in
Eq.~(\ref{eq:1stdef_SD_corr}) that originate from the
individual state $J$ are included into
the state-driven part of the correlation energy, thus leading to a second definition (that we
denote $\overline{\rm SD}$ to distinguish
the two decompositions): 
\be\label{eq:SD_a_la_Gould}
&&E^{{\bf w},{\rm
 \overline{SD}}}_{{\rm c},{J}}\left[n^{{\bf w}}\right]
:=
\mathcal{E}_{{\rm
c},J}^\bw\left[n^\bw\right]
\nonumber\\
&&+\tw_J\sum^{M}_{I=1}\left(\delta_{IJ}-{{\tt w}_I}\right)
\dfrac{\partial
}{\partial\tw_I}
\Big(\mathcal{E}_{{\rm
c},J}^{\bw}\left[n^{\bw}\right]
\Big),
\ee
which may be rewritten as~\cite{supp_mat} 
\be\label{eq:SDbar_written_like_Gould}
E^{{\bf w},{\rm
 \overline{SD}}}_{{\rm c},{J}}\left[n^{{\bf w}}\right]=
\bra{\Psi_J}\hat{T}+\hat{W}_{\rm ee}\ket{\Psi_J}-\overline{f}^{
\bw}_{J}\left[n^{{\bf w}}\right],
\ee
where the {\it effective} individual
exchange-only GOK energy reads 
\be\label{eq:eff_ind_X-only_ener}
\overline{f}^{
\bw}_{
J}\left[n^{{\bf w}}\right]
={f}^{
\bw}_{
J}\left[n^{{\bf w}}\right]+
{\tt w}_J\sum^{M}_{{I}=1}\left(\delta_{I{J}}-{{\tt
w}_I}\right)\dfrac{\partial {f}^{
\bw}_{
J}\left[n^{{\bf w}}\right]}{\partial \tw_I},
\ee
and the $J$th component of the exchange-only density-functional GOK energy can be
extracted from the ensemble one as follows: 
\be\label{eq:extraction_ind_X-only_ener}
f^\bw_J\left[n^{{\bf
w}}\right]=f^\bw\left[n^{{\bf
w}}\right]
+\sum^M_{I=1}\left(\delta_{IJ}-\tw_I\right)
\left.\frac{\partial
f^{\bxi}
\left[n^{\bxi,\bw}\right]
}{\partial \xi_I}
\right|_{\bxi=\bw}.
\ee
Equations~(\ref{eq:SDbar_written_like_Gould})--(\ref{eq:extraction_ind_X-only_ener})
are the second key result of this Letter. They clearly show that a
Gould--Pittalis-like~\cite{PRL19_Gould_DD_correlation} state-driven
correlation energy can be constructed 
without any additional state-specific KS system. The latter is somehow
implicitly defined in the present approach through the extraction
procedure described in Eqs.~(\ref{eq:eff_ind_X-only_ener}) and
(\ref{eq:extraction_ind_X-only_ener}), thus circumventing potential representability and non-uniqueness issues mentioned in the
introduction. Note that 
the first (SD) and second ($\overline{\rm SD}$) correlation energies can
be connected as
follows:
\be\label{eq:connection_SD-SDbar}
\begin{split}
&E^{{\bf w},{\rm
\overline{
SD}}}_{{\rm c},{J}}\left[n^{{\bf w}}\right]=
E^{{\bf w},{\rm
SD}}_{{\rm c},{J}}\left[n^{{\bf w}}\right]
\\
&+\sum^M_{K\neq J}\sum^{M}_{{I}=1}{\tt w}_K\left(\delta_{I{J}}-{{\tt
w}_I}\right)\frac{\partial {f}^{
\bw}_{
K}\left[n^{{\bf w}}\right]}{\partial \tw_I}.
\end{split}
\ee
Moreover, as both decompositions should
return the same individual correlation energy, the complementary
$\overline{\rm DD}$ part will simply be defined as 
$E^{{\bf w},{\rm
\overline{
DD}}}_{{\rm c},{J}}\left[n^{{\bf w}}\right]
:=
E^{{\bf w},{\rm
DD}}_{{\rm c},{J}}\left[n^{{\bf w}}\right]
-(E^{{\bf w},{\rm
\overline{
SD}}}_{{\rm c},{J}}\left[n^{{\bf w}}\right]
-E^{{\bf w},{\rm
SD}}_{{\rm c},{J}}\left[n^{{\bf w}}\right]
)$.\\

Let us now discuss the relevance of the two (in-principle-exact)
decompositions. Opting for one or the other will depend, in practical calculations, on
the level of approximation that is considered. If one is
able to construct a functional that incorporates weight dependencies (from a finite uniform electron gas~\cite{loos2020weightdependent}, for
example), then the first decomposition can be applied straightforwardly
in order to compute the energy levels (and properties) of the excited
states.
In the most common situation, where no weight-dependent functional is
available, the second decomposition might be preferred. One of the
reason is that individual DD correlation energies [see Eq.~(\ref{eq:alternative_exp_DD_corr})] read $\sum^M_{I=1}(\delta_{IJ}-\tw_I)\Delta_I$
and, since $\sum^M_{J=0}\sum^M_{I=1}\tw_J(\delta_{IJ}-\tw_I)\Delta_I=0$,
they are {\it
traceless}, i.e.
\be
\sum^M_{J=0}\tw_J\,E^{{\bf w},{\rm
DD}}_{{\rm c},{J}}\left[n^{{\bf w}}\right]
=
0.
\ee
In other words, DD correlations contribute to the  
individual energies, not to the ensemble one.  
This statement holds at any level of approximation, by construction. As
a result, the (first) SD/DD decomposition cannot be used for developing
density-functional approximations to the ensemble
correlation energy. As further discussed in the following, the (second)
$\overline{\rm SD}$/$\overline{\rm DD}$ decomposition is more appealing in
this respect, simply because $\overline{\rm DD}$ correlations
are {\it not} traceless. As shown in the supplementary
material~\cite{supp_mat}, the total $\overline{\rm DD}$ ensemble
correlation energy can be evaluated exactly from the linear response of
the exchange-only GOK energy
components as follows: 
\be\label{eq:final_exp_ens_DDbar_corr}
E^{{\bf w},{\rm
\overline{DD}}}_{{\rm c}}\left[n^{{\bf w}}\right]
=\sum^M_{J=0}{\tt w}^2_J
\sum^{M}_{{I}=1}\left(\delta_{I{J}}-{{\tt
w}_I}\right)\frac{\partial {f}^{
\bw}_{
J}\left[n^{{\bf w}}\right]}{\partial \tw_I}
.
\ee

{\it Approximations.} The most challenging task in
GOK-DFT, namely the design of correlation 
density functionals for ensembles, can now be addressed by modeling 
$\overline{\rm SD}$ and $\overline{\rm DD}$ ensemble correlations separately.
For the former, one may rely on the conventional ground-state limit of
GOK-DFT, i.e. $E^{{\bf w},{\rm
\overline{SD}}}_{{\rm c},{J}}\left[n^{{\bf w}}\right]\approx E^{\bw={\bf 0},{\rm
\overline{SD}}}_{{\rm c},J}\left[n_{\Psi_0}\right]$, 
where, according to Eqs.~(\ref{eq:SDbar_written_like_Gould}) and (\ref{eq:eff_ind_X-only_ener}), $E^{\bw={\bf 0},{\rm
\overline{SD}}}_{{\rm c},J=0}\left[n_{\Psi_0}\right]=E_{\rm
c}\left[n_{\Psi_0}\right]$ is the conventional ground-state
density-functional correlation
energy and 
$E^{\bw={\bf 0},{\rm
\overline{SD}}}_{{\rm c},J}\left[n_{\Psi_0}\right]\overset{J>0}{=}
\bra{\Psi_J}\hat{T}+\hat{W}_{\rm ee}\ket{\Psi_J}
-\bra{\Phi_J}\hat{T}+\hat{W}_{\rm ee}\ket{\Phi_J}$, 
$\Phi_J$ being an excited KS wavefunction based on a regular DFT
calculation. Starting from the crudest {\it ground-state-correlation}
only (GSc-$\overline{\rm SD}$) weight-dependent approximation, 
$E^{{\bf w},{\rm \overline{SD}}}_{{\rm c}}\left[n^{{\bf
w}}\right]\overset{{\rm GSc}-\overline{\rm SD}}{\approx}
(1-\sum^M_{J=1}\tw_J)E_{\rm
c}\left[n_{\Psi_0}\right]$,
correlation in the excited states might be introduced in various ways,
as explored in a recent work by
Gould and Pittalis~\cite{Gould_2020_insights}. One may also learn either from models such as finite electron
gases~\cite{loos2020weightdependent} or even from
TD-DFT excitation energies $\omega_J$ through the following exact
relation, 
\be
&&E^{\bw={\bf 0},{\rm
\overline{SD}}}_{{\rm
c},J}\left[n_{\Psi_0}\right]\overset{J>0}{=}E_0+\omega_J-\langle\Phi_J\vert\hat{H}\vert\Phi_J\rangle
\nonumber
\\
&&-\int
d\br\,v_{\rm ext}(\br)\Big(
n_{\Psi_J}(\br)-n_{\Phi_J}(\br)
\Big),
\ee
and Eq.~(\ref{eq:deviation_true-KS_dens}), for the evaluation of the
density difference.
As GSc-$\overline{\rm
SD}$ turned out to be accurate enough for the Hubbard dimer (see
the next section), we leave the design of approximations beyond GSc-$\overline{\rm 
SD}$ for future work. On the other hand, the $\overline{\rm DD}$
part in Eq.~(\ref{eq:final_exp_ens_DDbar_corr}) can be evaluated accurately through finite differences with optimized effective potential
techniques~\cite{yang2017direct}. 
We propose here a simpler {\it ground-state-exchange} functional approximation
(GSx-$\overline{\rm DD}$), 
$f^{\bxi}[n^{\bxi,\bw}]\overset{{\rm GSx}-\overline{\rm
DD}}{\approx}\sum^M_{K=0}\xi_K\bra{\Phi_K^\bw}\hat{T}\ket{\Phi_K^\bw}+E_{\rm
Hx}[n^{\bxi,\bw}]$, which is tested in the following.\\
 
{\it Application.} We will now explore the two decompositions in the
asymmetric two-electron Hubbard
dimer model~\cite{carrascal2015hubbard,deur2017exact,deur2018exploring,sagredo2018can,senjean2018unified,deur2019ground}
where, due to representability issues mentioned previously, the
Gould--Pittalis decomposition cannot
be applied. The Hubbard dimer can be seen as a prototype for a diatomic
molecule where the density $n$ reduces to a (possibly
fractional) number that corresponds to the occupation of the first
atomic site
[the occupation of the second atom is then $2-n$]. It is governed by
three parameters: the hopping $t$ that modulates the strength of the
kinetic energy, the on-site two-electron repulsion strength $U$, and the
external potential difference $\Delta v_{\rm ext}$ which controls the
asymmetry of the dimer. 
For simplicity, we focus on the weakly
asymmetric and strongly correlated regime 
$\Delta v_{\rm ext}/t<<t/U<<1$. In this case,
the singlet biensemble density
reads~\cite{deur2017exact,deur2018exploring,supp_mat}
$n^\tw\approx1+\tw\eta$,
where $\tw\equiv\tw_1$ and $\eta=(U\Delta v_{\rm
ext})/(2t^2)<<1$. As shown in the supplementary material~\cite{supp_mat}, each individual
correlation energy (and the subsequent decompositions) can be derived
analytically. For example, for the 
excited state (whose charge-transfer character
increases with $\Delta v_{\rm ext}/t$), we obtain the following expressions [we denote
$\mathscr{E}\equiv E/(U\eta^2)$ energies per unit of $U\eta^2$]:
\be\label{eq:SD-DD_corr_ener_ES}
&\mathscr{E}^{\tw,\rm SD}_{{\rm c},J=1}\left(n^\tw\right)
\approx
\dfrac{\tw(4\tw-1)}{(1-\tw)^2},
\mathscr{E}^{\tw,\overline{\rm SD}}_{{\rm c},J=1}\left(n^\tw\right)\approx 
\dfrac{3\tw^2}{(1-\tw)^2} 
\\
\label{eq:SD-DDbar_corr_ener_ES}
&\mathscr{E}^{\tw,\rm DD}_{{\rm c},J=1}\left(n^\tw\right)
\approx
\dfrac{\tw(1-3\tw)}{(1-\tw)^2},\mathscr{E}^{\tw,\overline{\rm
DD}}_{{\rm c},J=1}\left(n^\tw\right)\approx-\dfrac{2\tw^2}{(1-\tw)^2}. 
\ee
As readily seen from
Eqs.~(\ref{eq:SD-DD_corr_ener_ES}) and (\ref{eq:SD-DDbar_corr_ener_ES}), 
individual correlation energies within an ensemble can be positive. This is not surprizing as
the energy extraction procedure used in Eq.~(\ref{eq:ind_ener_from_KS})
is {\it not} variational,
even though the ensemble energy is. We also clearly see that, in both
decompositions,
density-driven correlations can be substantial. In the
equiensemble ($\tw=1/2$) case, the 
density-/state-driven correlation energy ratio equals 50\% and 66\% in
the first and second decompositions, respectively. 
We essentially reach the same conclusions as Gould and Pittalis~\cite{PRL19_Gould_DD_correlation}, even though 
we use a different decomposition. Turning to the $\overline{\rm SD}$/$\overline{\rm DD}$
decomposition of the ensemble correlation energy, the following
expression is obtained in the regime under study~\cite{supp_mat}, 
\be\label{eq:ens_corr_SD}
\mathscr{E}^{\tw,\overline{\rm SD}}_{{\rm c}}\left(n^\tw\right)&\approx&
-\dfrac{1}{2\eta^2}(1-\tw)+
\dfrac{\tw^2(1+5\tw)}{2(1-\tw)^2},
\\
\label{eq:ens_corr_DD}
\mathscr{E}^{\tw,\overline{\rm DD}}_{{\rm c}}\left(n^\tw\right)&\approx&
-\dfrac{\tw^2(1+\tw)}{(1-\tw)^2},
\ee
while the density-functional approximations introduced previously give:
\be\label{eq:ens_corr_SD_approx}
\mathscr{E}^{\tw,\overline{\rm SD}}_{{\rm
c}}\left(n^\tw\right)&\overset{{\rm GSc}-\overline{\rm SD}}{\approx}&
-\dfrac{1}{2\eta^2}(1-\tw),
\\
\label{eq:ens_corr_DD_approx}
\mathscr{E}^{\tw,\overline{\rm DD}}_{{\rm
c}}\left(n^\tw\right)&\overset{{\rm GSx}-\overline{\rm
DD}}{\approx}&-\tw^2\left[1+2\tw(1-\tw)\right].
\ee
As we used exact densities, density-driven 
errors~\cite{deur2018exploring} have been neglected.
Their study is left for future work. 
Around the leading (second) order in $1/\eta>>1$, which is correctly
reproduced within the simple GSc-$\overline{\rm
SD}$ approximation, both $\overline{\rm
SD}$ and $\overline{\rm
DD}$ correlations contribute (almost equally in the equiensemble case)
to the ensemble correlation energy, but with
opposite sign, thus reducing substantially the weight dependence [see Eqs.~(\ref{eq:ens_corr_SD}) and
(\ref{eq:ens_corr_DD})]. While an accurate description of the $\overline{\rm
DD}$ correlation on top of GSc-$\overline{\rm
SD}$ overestimates the total correlation energy,
a drastic improvement is obtained by combining GSx-$\overline{\rm DD}$
with GSc-$\overline{\rm
SD}$, thanks to error cancellations [see the supplementary material~\cite{supp_mat} for further
details]. This promising result might be
further improved by incorporating $\overline{\rm
SD}$ correlations from the excited state, which is left for future work. 
\\

{\it Summary and outlook.} 
By uncovering the individual correlation energies within a density-functional
ensemble, we were able to derive a
state-/density-driven decomposition of the ensemble correlation where,
unlike in the Gould-Pittalis
decomposition~\cite{PRL19_Gould_DD_correlation}, 
no additional state-specific KS system is needed. 
By expressing the density-driven ensemble correlation energy  
in terms of the ensemble weights, the ensemble KS orbitals, and their
(static) linear response to variations in
the weights, we made a crucial step
toward the development of first-principle density functionals for
ensembles. While we focused on 
individual energies,
the extension of the theory to energy couplings such as
transition dipole moments or non-adiabatic
couplings is highly desirable. 
Work is currently in progress in these directions.\\

{\it Acknowledgments.} The author would like to thank Pierre-Fran{\c c}ois
Loos and Bruno Senjean for fruitful discussions, and LabEx CSC (ANR-10-LABX-0026-CSC)
for funding.




\newcommand{\Aa}[0]{Aa}

\end{document}